\documentclass[pra,twocolumn,showpacs,preprintnumbers,amsmath,amssymb,floatfix]{revtex4}
\usepackage{epsfig}
\usepackage{graphicx}
\usepackage{amssymb}
\usepackage{color}

\begin{document}

\title{Entanglement Storage Units}

\author{Tommaso Caneva$^{1}$}
\author{Tommaso Calarco$^1$}
\author{Simone Montangero$^{1}$}

\address{$^1$Institut f\"ur Quanteninformationsverarbeitung,
  Universit\"at Ulm, D-89069 Ulm, Germany}
%\ead{tommaso.caneva@uni-ulm.de}

\date{\today}

\begin{abstract}
We introduce a protocol based on optimal control to drive many body quantum systems 
into long-lived entangled states, protected from decoherence by big
energy gaps, without requiring any apriori knowledge of the system. 
With this approach it is possible to implement scalable 
entanglement-storage units. We test the protocol in the Lipkin-Meshkov-Glick 
model, a prototype  many-body quantum system that
describes different experimental setups, and in the ordered  
Ising chain, a model representing a possible implementation of a quantum
bus.
%undergoing a second order quantum phase transition.
%
\end{abstract}


\maketitle

%%%%%%%%%%%%%%%%%%%%%%%%%%%%%%%%%%%%%%%%%%%%
%\section{Introduction}    %%%%%%%%%%%%%%%%%
%%%%%%%%%%%%%%%%%%%%%%%%%%%%%%%%%%%%%%%%%%%%
\section{Introduction}
 
Entanglement represents the manifestation of correlations
without a classical counterpart and it is regarded as the necessary ingredient
at the basis of the power of quantum information processing. Indeed
quantum information applications as teleportation, quantum cryptography or quantum 
computers rely on entanglement as a crucial resource~\cite{Nielsen_Chuang:book}.
Within the current state-of-art, promising candidates for truly scalable
quantum information processors are considered architectures that interface hardware components 
playing different roles like for example solid-state systems as stationary 
qubits combined in hybrid architectures with optical devices~\cite{Baumann_NAT10}. 
In this scenario, the stationary qubits are a collection of engineered qubits with 
desired properties, as decoupled as possible from one another to prevent errors. 
However, this architecture is somehow unfavorable to the creation and the conservation of entanglement. 
Indeed, it would be desirable to have a hardware  where ``naturally'' 
entanglement is present and that can be prepared in 
a highly entangled state that persists without any external control: 
the closest quantum entanglement analogue of a classical information 
memory support, i.e. an \emph{entanglement-storage unit} (ESU). 
Such hardware once prepared  can be used at later times
(alone or with duplicates) -- once the desired kind of
entanglement has been distilled -- to perform quantum information 
protocols~\cite{Nielsen_Chuang:book}. 

The biggest challenge in the development of an ESU is entanglement 
frailty: it is strongly affected by the detrimental presence of decoherence~\cite{Nielsen_Chuang:book}.
Furthermore the search for a proper system to build an ESU is undermined by 
the increasing complexity of quantum systems with a growing
number of components, which makes entanglement more frail, 
more difficult to characterize, to create and to control~\cite{Amico_RMP08}. 
Moreover, given a many body quantum system, the search for a state with the desired
properties 
%CUT might be very difficult. Indeed, a direct and comprehensive study of a many body quantum system 
is an exponentially hard task in the system size.
Nevertheless, in many-body quantum
systems entanglement naturally arises: for example --when undergoing a
quantum phase transition -- in proximity of a critical point the
amount of entanglement possessed by the ground state scales with the 
size~\cite{Amico_RMP08,Vidal1_PRL03}. Unfortunately, due
to the closure of the energy gap at the critical point, the ground state 
is an extremely frail state: even very little perturbations might
destroy it, inducing excitations towards other states.
However a different strategy might be successful, corroborated also by very
recent investigations on the entanglement properties of the eigenstates of many-body 
Hamiltonians, where it has been shown
that in some cases they are characterized by entanglement growing with the system 
size~\cite{Alba_JSTAT09,Alcaraz_PRL11}. 

\begin{figure}
\centering\epsfig{file=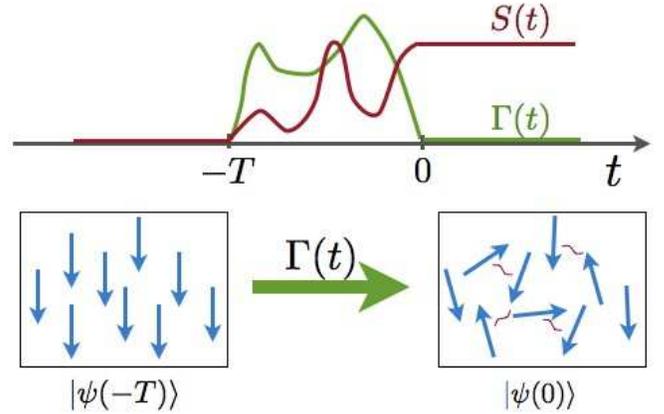,width=9cm,angle=0,clip=}
\caption{(Color online) Entanglement Storage Units protocol: a system
  is initially in a reference state $|\psi(-T)\rangle$, e.g. the ground
  state, and is optimally driven via a control field $\Gamma(t)$ 
  in an entangled eigenstate $|\psi(0)\rangle$, 
  protected from decoherence by an energy gap. $S(t)$ represents
  a generic measure of entanglement.}
\label{scheme}
\end{figure}
In this work we show that by means of
a recently developed optimal control technique~\cite{Doria_PRL11,Caneva_PRA11} 
it is possible to identify and prepare a many body quantum system in robust, long-lived entangled states 
(ESU states). More importantly, we drive the system towards ESU states without the need of any apriori 
information on the system, either about the eigenstates or about the energy spectrum. 
%CUT Finally, we show that properly prepared systems can be effectively used as ESU exploiting 
%the fact that ESU states are well protected by large energy gaps. 
%
Indeed, we do not first solve the complete spectrum and eigenstates, which is 
an exponentially difficult problem in the system size.
Recently, optimal control has been used to drive quantum systems in 
entangled states or to improve the generation of
entanglement~\cite{Platzer_PRL10}. However, here we have 
in mind a different scenario: to exploit the control
to steer a system into a highly entangled state that is stable and
robust even after switching off the control (see Fig.~\ref{scheme}). 
Moreover we want to outline the fact that we do not choose
the goal state, but only its properties.
In the following we show that ESU states
are gap-protected entangled eigenstates of the system 
Hamiltonian \emph{in the absence} of the 
%CUT control. 
%Here we show that for an experimentally relevant model this is indeed
%possible, and that it is possible to drive the system in gap-protected
%states. 
control, and that for an experimentally relevant model is indeed 
possible to identify and drive the system into the ESU states.
We show that the ESU states, although not 
being characterized by the maximal entanglement sustainable by
the system, are characterized by entanglement that grows
with the system size. 
Once a good ESU state has been detected, due to
its robustness it can be stored, characterized, and thus used for later quantum 
information processing.\\ 
Here we provide an important example of this approach, based on the 
the Lipkin-Meshkov-Glick (LMG) model~\cite{Lipkin_NP65}, a system  
realizable in different experimental setups~\cite{Baumann_NAT10,Buecker_NAT11}; we prepare an
ESU maximizing the Von Neumann entropy of a bipartition of the system
and we model the action of the surrounding environment with noise terms
in the Hamiltonian. However, our protocol is compatible 
with different entanglement measures and different models, 
like
the concurrence between the extremal spins in an Ising chain,
see Sec.~\ref{ising:sec}. Notice that
with a straightforward generalization it can be adapted to a 
full description of open quantum systems~\cite{inpreparation}. 

The article is organized as follows: in Sec.~\ref{protocol:sec} 
the general protocol to steer a system onto ESU state is presented; 
in Sec.~\ref{model:sec} we consider the application of the protocol
to the Lipkin-Meshkov-Glick model; in 
Sec.~\ref{noise:sec} we discuss the effect of a telegraphic classical noise 
onto the protocol; in Sec.~\ref{ising:sec} we test the protocol 
into an Ising spin chain,
and finally in Sec.~\ref{conclusion:sec} we present the 
conclusions of our work.
\begin{figure}
\centering\epsfig{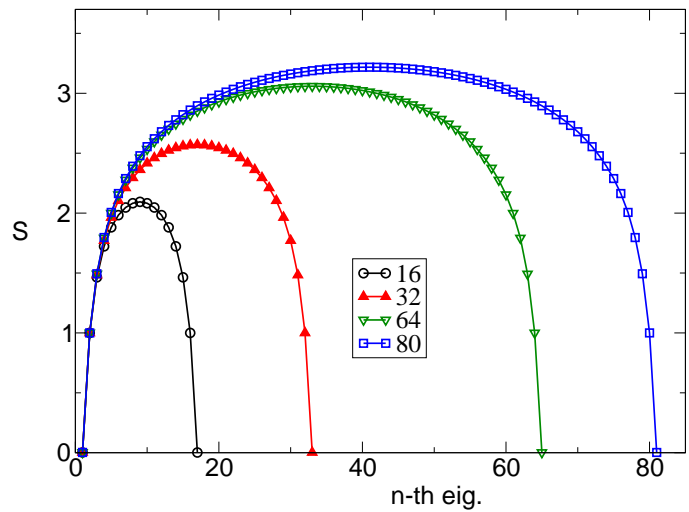} %height=4.8cm
\caption{(Color online) LMG model: Static entanglement $S$ of the
eigenstates at $\tilde{\Gamma}=10$ for different system sizes $N=16,32,64,80$. The eigenstates
are ordered according to their energy, i.e. $n=1$ corresponds to the ground state.}
\label{lmg_central_eig_ent_vs_size_1:fig}
\end{figure}

\section{ESU protocol}
\label{protocol:sec} 
As depicted in Fig.~\ref{scheme}, we consider the general scenario of
a system represented by a tunable Hamiltonian 
%$H_o$ with an additional tunable term  $H_1[\Gamma]$
 $H[\Gamma]$, where $\Gamma(t)$ is the control field, and
initialized in a state $|\psi _{in}\rangle$ that can be easily prepared. 
%In the following we are assuming the common situation in which the initial
%state corresponds to the ground state of $H$ at a particular
%value of the field $\tilde{\Gamma}$, e.g. $|\psi _{in}\rangle =|\tilde{H}_{gs} \rangle$, 
%with $\tilde{H}=H[\tilde{\Gamma}]$.
%%The Hamiltonian is then $H=H_0+H_1[\Gamma]$, where the
%%control parameter is initially set at a constant value (in particular it can
%%vanish, $\Gamma(0)=0$). 
We assume that the control field $\Gamma(t)$ can be modulated only in the finite 
time interval $[-T,0]$; outside of this interval, for $t<-T$ and $0<t$, we impose
$\Gamma \equiv \tilde{\Gamma}$ (e.g. absence of control).
According to our protocol, at the end of the control procedure, i.e. once the 
control field is
brought back to the value  $\tilde{\Gamma}$, the system has been prepared 
in a state with desired properties (for instance high entanglement), 
stable in absence of the control and robust against noise and perturbations.\\
Optimal control has been already used to enhance a given desired property without
targeting an apriori known state; unfortunately the results of such optimization
are usually fragile and ideally require a continuous application of the control in order to be
stabilized~\cite{Platzer_PRL10}. However in practical situations a continuous application 
of control can be unrealistic, 
being either simply impossible or too expensive in terms of resources.
An example is the initialization of a quantum register that 
has to be physically moved into different spatial locations (like a portable memory support),
or if the control field used to initialize has to be switched on and off
in order to 
manipulate different parts of the apparatus; in such situations indeed 
the register should be stable also once disconnected from 
the device employed for its initialization.
Consequently in certain applications, a procedure capable to prepare quantum targets 
intrinsically stable even in the absence of sustained external manipulations is not only 
highly desirable 
but also crucial. The main contribution of our work is exactly to move a step forward
in this direction, proposing a flexible 
recipe to improve the stability of the outcome of a generic optimization process.\\
The simply idea behind our method is the following: as it is well known, in a closed system, the 
evolution of an arbitrary state is driven by Schr\"odinger equation 
$i\hbar |\dot{\psi}(t)\rangle=H(t)|\psi(t)\rangle$. Assuming that, as in the absence of control, 
the Hamiltonian is constant $H(t)= H[\tilde{\Gamma}]$, we can evaluate the extent of 
the deviation induced by the time evolution in an infinitesimal time $dt$ after switching off
the control~\cite{Anandan_PRL90}:
\begin{eqnarray}
 1-|\langle \psi(t)|\psi (t+dt)\rangle |^2=\Delta \tilde{E} ^2 dt ^2/\hbar ^2 + O(dt ^3),
\label{psi_ev:eq}
\end{eqnarray}
where $\Delta \tilde{E}=\sqrt{\langle\psi (t)|H^2[\tilde{\Gamma}]|\psi (t)\rangle -\tilde{E} ^2}$ and 
$ \tilde{E}=\langle\psi (t)|H[\tilde{\Gamma}]|\psi (t)\rangle$ correspond respectively to
the energy fluctuations and the energy of the Hamiltonian in absence of control.
Then from Eq.~(\ref{psi_ev:eq}) it is clear that an arbitrary state is stabilized by
minimizing the quantity $\Delta \tilde{E}$. In particular, by reaching the condition
$\Delta \tilde{E}=0$, the system is also prepared in an eigenstate of
$H[\tilde{\Gamma}]$.\\
Our protocol relies on the use of optimal control implemented through 
the Chopped RAndom Basis (CRAB) technique\cite{Doria_PRL11,Caneva_PRA11}. The
CRAB method consists in expanding the control field
onto a truncated basis (e.g. a truncated Fourier series) and
in minimizing an appropriate cost function with respect to the weights
of each component of the chopped basis (see~\cite{Doria_PRL11,Caneva_PRA11} for details of 
the method).\\
In particular, for the ESU protocol a CRAB optimization is performed 
%for $t\in [-T,0]$ 
with the goal of minimizing the cost function 
$\mathcal{F}$:
\begin{eqnarray}
\mathcal{F}(\lambda)_{|\psi (0)\rangle}=-S+ \lambda \frac{\Delta \tilde{E}}{\tilde{E}},
\label{cost_f:eq}
\end{eqnarray}
where $S$ represents a measure of entanglement, 
%$\Delta \tilde{E}=\sqrt{\langle\psi|H^2[\tilde{\Gamma}]-\langle H[\tilde{\Gamma}]\rangle ^2|\psi\rangle}$ and 
%$ \tilde{E}=\langle\psi|H[\tilde{\Gamma}]|\psi\rangle$ correspond respectively to
%the energy fluctuations and the energy of the Hamiltonian in absence of control, 
$\lambda$ is a Lagrange multiplier, and the cost function is evaluated 
on the optimized evolved state $|\psi (0)\rangle$ produced with a control 
process active in the time interval $[-T,0]$.
As discussed previously and shown in the following, the inclusion in $\mathcal{F}$ of the
constraint on the energy fluctuations is the crucial 
ingredient to stabilize the result of the optimization also for times $t>0$, that is
once the control has been switched off.\\ 
We conclude this section stressing a couple of important advantages of our 
protocol with respect to possible other approaches to the problem, like
for instance evaluating all the eigenstates of the system and picking up 
among them the state(s) with the desired properties. First, in our
protocol we never compute the whole spectrum of the system, but we simply require
to evaluate the energy and the energy fluctuations into the evolved state,
see Eq.~(\ref{cost_f:eq}); therefore our procedure can be applied also
to situations in which it is not possible to compute all the 
eigenstates of the Hamiltonian (e.g. many-body non integrable systems or DMRG simulations
or experiments including a feedback loop).
Furthermore it can occur that none of the eigenstates of the system owns the desired
property we would like to enhance; then by simply considering the eigenstates 
one could not gain any advantage. On the contrary, also in this situation, with our protocol it is 
possible to identify states that, even though different from exact eigenstates, anyway show an 
enhanced robustness, like the optimal state found in the considered scenario, 
see Sec.~\ref{ising:sec}.
\begin{figure}
\centering\epsfig{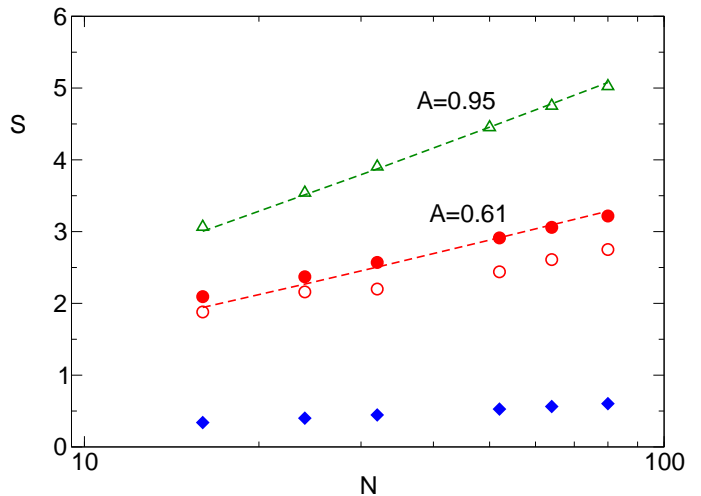} %height=4.8cm
\caption{(Color online) LMG model: Ground state entanglement at the 
critical point of the (full blue diamonds); 
central eigenstate entanglement at $\tilde{\Gamma}=10$ (full red circles); 
maximal eigenstate entanglement obtained with the optimization for
$\lambda\neq 0$ (empty red circles) and $\lambda = 0$ (green triangle).
The red (green) dashed line is numerical fit $A\cdot \log _2 (N/2+1)$
with $A=0.61$ ($A=0.95$).}
\label{lmg_central_eig_ent_vs_size_2:fig}
\end{figure}
\section{ESU and Lipkin-Meshkov-Glick model} 
\label{model:sec}

% Here we provide an important example of this approach, based on the 
% the Lipkin-Meshkov-Glick (LMG) model~\cite{Lipkin_NP65}; we prepare an
% ESU maximizing the Von Neumann entropy of a bipartition of the system
% and we model the action of the surrounding environment with noise terms
% in the Hamiltonian. However, our protocol is compatible 
% with different entanglement measures and different models, 
% and with a straightforward generalization it can be adapted to a full description of 
% open quantum systems~\cite{inpreparation}.  
We decided to apply the protocol to the Lipkin-Meshkov-Glick model\cite{Lipkin_NP65}
because it represents an interesting prototype of the challenge we address: it
describes different experimental setups~\cite{Baumann_NAT10,Buecker_NAT11},
and the entanglement properties of the eigenstates are in general not known. 
Indeed, %the LMG model has been recently investigated in different 
%experimental setups~\cite{Baumann_NAT10,Buecker_NAT11}.
the entanglement properties of the eigenstates of one-dimensional
many-body quantum systems have been related with 
the corresponding conformal field theories~\cite{Alba_JSTAT09}; however for the LMG model,
to our knowledge, this study has never been performed and a conformal theory is not 
available~\cite{Latorre_JPA09}. Finally, the optimal control problem
we address is highly non-trivial as the control field is global and
space-independent with no single-site addressability~\cite{Platzer_PRL10}.

The LMG Hamiltonian describes an ensemble of spins with infinite-range
interaction and is written as\cite{Botet_PRB83}:
\begin{eqnarray}
H=-\frac{C}{N}\sum ^N _{i< j}\sigma _i ^x \sigma _j ^x -
    \Gamma (t) \sum _{i}^N \sigma _i ^z ,
\label{full_ham_lmg:eq}
\end{eqnarray}
where $N$ is the total number of spins, $\sigma _i^{\alpha}$'s ($\alpha =x,y,z$ ) are the Pauli matrices on 
the $i$th site and $C$ is a constant measuring the intensity of the spin-spin interaction. 
%CUT and $J_{ij}=J/N$ (infinite range interaction). 
By introducing the total spin operator $\vec{J}=\sum_i \vec{\sigma}_i/2$, 
the Hamiltonian can be rewritten, apart from an additive constant and a constant 
factor as
\begin{eqnarray}
H=-\frac{1}{N}J_x^2-\Gamma J_z,
\label{ham_lmg:eq}
\end{eqnarray}
\begin{figure}[t]
\centering\epsfig{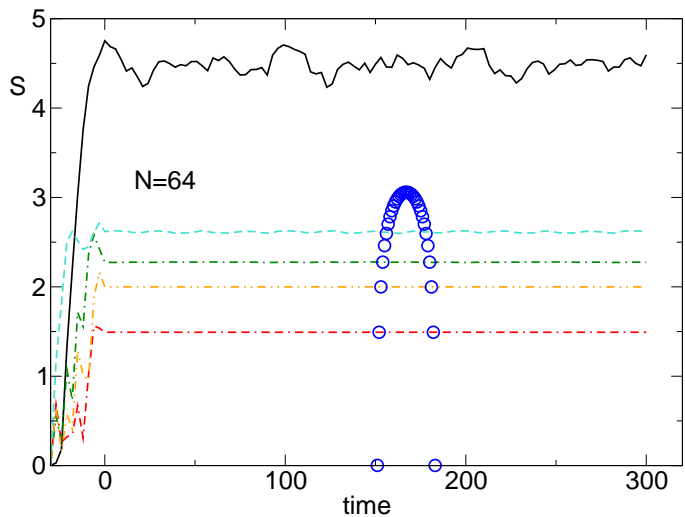}
\caption{(Color online) Entanglement entropy $S(t)$ as a function of
  time (time unit $C^{-1}$) for different $\lambda$ values: $\lambda=0$  (black) continuous, $\lambda=5$  (red) dash-dash-dotted, 
$\lambda=1.8$ (green) dot-dashed, $\lambda=1.9$ (orange) dot-dot-dashed, $\lambda=1.2$  (cyan) dashed line. 
Blue circles represent the entropy of the eigenstates for $N=64$ and $\tilde{\Gamma} =10$.}
\label{ent_inf_vs_timeWL64_1:fig}
\end{figure}
(from now on we set $C=1$ and $\hbar=1$).
%CUT The Hamiltonian hence commutes with $J^2$ and does not couple states having a
%different parity  of the number of spins pointing in the magnetic field
%direction: $[H,J^2]=0$ and $[H, \prod _i \sigma _i ^z]=0$. 
The symmetries of the Hamiltonian imply that the dynamics is restricted to 
subspaces of fixed total magnetization $J$ 
and fixed parity of the projection $J_z$;
a convenient basis 
for such subspaces is represented by the Dicke states $|J,J_z\rangle$ with
$-J<J_z<J$~\cite{Latorre_PRA05}. 
In the thermodynamical limit the system undergoes a 2nd order QPT
from a quantum paramagnet to a quantum ferromagnet
at a critical value of the transverse field $|\Gamma_ c| =1$.
%The gap between the ground state and first
%excited state closes polynomially with the size at the critical point: $\Delta _{\rm LMG}\sim N^{-1/3}$. 
There is no restriction to the 
 reference value $\tilde{\Gamma}$
%CUT (in the implementation of Ref.~\cite{Baumann_NAT10} it goes to infinity when the
%control lasers are switched off) 
and to the initial state $|\psi_{in}\rangle$: 
we choose $\tilde{\Gamma}\gg 1$, corresponding to the paramagnetic phase
%~\footnote{Indeed the structure of the Hamiltonian suggests that in the paramagnetic phase
%the typical energy scale is proportional to the transverse field $\Gamma$; in the ordered phase
%instead such a scale is proportional to $1/N$. Consequently the energy density of states is
%expected to be higher in the ordered phase and an enhanced energy level separation should occur
%in the paramagnetic phase.},
and as initial state $|\psi _{in}\rangle$, the ground state of $H[\tilde{\Gamma}]$,
i.e. the separable state in which all the spins
are polarized along the positive $z$-axis~\cite{Baumann_NAT10}. %The GS belongs to the subspace with
%maximal angular momentum $J=N/2$, so that in our analysis we work in this subspace.
%As target state has been chosen the GS at $\Gamma=0$.\\
A convenient measure of the entanglement in the LMG model is given by
the von Neumann entropy
$S_{L,N}=-{\rm Tr}(\rho _{L,N}\log _2\rho _{L,N})$ associated
to the reduced density matrix $\rho _{L,N}$ of a block of $L$ spins
out of the total number $N$, which gives a
measure of the entanglement present between two partitions of a
quantum system~\cite{Latorre_PRA05}. 
In our analysis we consider two equal partitions, i.e. $S\equiv S_{N/2,N}$.
Note that the maximally entangled state at a fixed 
size $N$ is given by $\rho _M={\bf 1}/(N/2+1)$ and $S_{\rho _M}=\log
_2(N/2+1)$ \cite{Latorre_PRA05}.
%(see Fig.~\ref{lmg_central_eig_ent_vs_siz_fig} and Ref).
%%%%%%%%%%%%%%%%%%%%%%%%%%
%\emph{Statics.}---
%%%%%%%%%%%%%%%%%%%%%%%%%%
In Fig.~\ref{lmg_central_eig_ent_vs_size_1:fig} we report
the entanglement $S_{N/2,N}$ of the eigenstates deeply inside the paramagnetic phase at $\tilde{\Gamma} =10$,
for systems of different sizes. Clearly, also far from the critical
point $\Gamma=1$ many eigenstates possess a remarkable amount of
entanglement that scales with the system size. 
%
%\begin{figure}
%\epsfig{file=plot/lmg_eigenstate_static_ent.eps,width=7cm,angle=0,clip=}
%\caption{(Color online) Static entanglement of the eigenstates at $\Gamma=10$ for different sizes. The eigenstates
%are ordered according to their energy, i.e. $n=1$ corresponds to the ground state.}
%\label{lmg_eigenstate_static_ent_fig}
%\end{figure}
%
%
%
The effect is shown more clearly in Fig.~\ref{lmg_central_eig_ent_vs_size_2:fig}, 
where the entanglement 
of the central eigenstate (red full circles) at $\tilde{\Gamma}=10$ is compared with the entanglement of the 
ground state
at the critical point (full blue diamonds).
%(full blue diamonds, see also Ref.~\cite{Barthel_PRL06}). 
Both sets of data show a logarithmic scaling with the size, but the entanglement of the central eigenstate is 
systematically higher and grows more rapidly.
\begin{figure}
\centering\epsfig{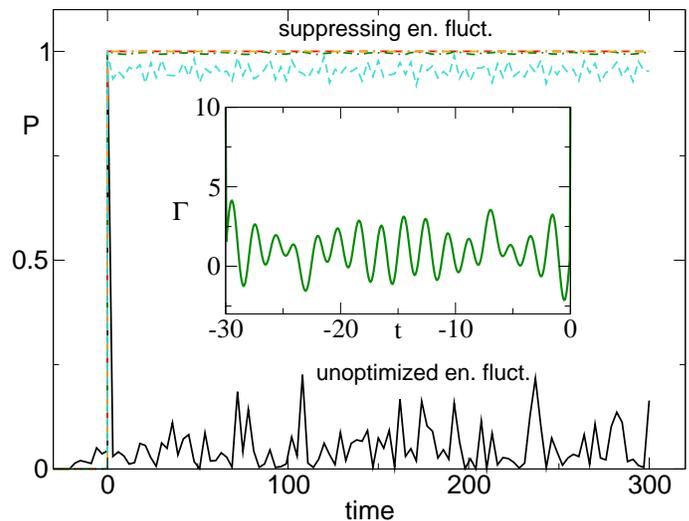}
\caption{(Color online) Survival probability $P(t)$ as a function of
time  (time unit $C^{-1}$) for different $\lambda$ values: $\lambda=0$  (black) continuous, $\lambda=5$  (red) dash-dash-dotted, 
$\lambda=1.8$ (green) dot-dashed, $\lambda=1.9$ (orange) dot-dot-dashed, $\lambda=1.2$  (cyan) dashed line. 
Inset: Optimal driving 
field $\Gamma(t)$ for $\lambda=1.8$ and $N=64$.}
\label{ent_inf_vs_timeWL64_2:fig}
\end{figure}
%
%
%%%%%%%%%%%%%%%%%%%%%%%%%%
\emph{Dynamics.}---
%%%%%%%%%%%%%%%%%%%%%%%%%%
We initialize the system in the non-entangled ground state of the Hamiltonian
%$H=H_0+H_1(\Gamma_0)$ 
$H[\tilde{\Gamma}]$ with $1\ll \tilde{\Gamma}=10$
so that in the absence of control, i.e. for
$\Gamma \equiv \tilde{\Gamma} $ independent 
of time, the state $|\psi_{in}\rangle$ does not evolve apart from a phase factor.
After the action of the CRAB-optimized driving field $\Gamma(t)$
for $t\in [-T,0]$ the state is prepared in $|\psi(0)\rangle$ (a typical 
optimal pulse is shown in the inset of Fig. ~\ref{ent_inf_vs_timeWL64_2:fig}), and
we observe the evolution of the state over times $t>0$. 
The behavior of the entanglement is shown in %Fig.~\ref{ent_inf_vs_timeWL16_fig} and 
Fig.~\ref{ent_inf_vs_timeWL64_1:fig} for different values of the
weighing factor $\lambda$ and $N=64$.
%CUT ; the control is active
%for negative times, i.e., in the interval $[-T,0]$. 
For $\lambda=0$ highly entangled states are produced, however the
entanglement $S(t)$ oscillates indefinitely with the time.
%CUT, reflecting
%the fact that the system state is changing over time.  
On the contrary, if the energy fluctuations are included in the cost
function ($\lambda\neq 0$),
the optimal driving field steers the system into 
entangled eigenstates of $H[\tilde{\Gamma}]$, 
as confirmed by the absence of the oscillations in the 
entanglement and by the entanglement 
eigenstate reference values (empty blue circles). These results are
confirmed by the survival probability in the initial state 
$P(t)=|\langle\psi(0)|\psi (t)\rangle|^2$ reported in 
Fig.~\ref{ent_inf_vs_timeWL64_2:fig}: the state prepared with
$\lambda=0$ decays over very fast time scales $\tau_0$, 
while for $\lambda\neq 0$ it 
remains close to the unity for very long times $\tau_\lambda >> \tau_0$. 
The small residual oscillations for $N=64$ and $\lambda=1.2$
are due to the fact that in this case the optimization leads to a state corresponding to
an eigenstate up to $98\%$. We repeated the optimal preparation for
different system sizes and initial states, and show the entanglement of the optimized
states for $\lambda=0$ (empty green triangles) and  $\lambda\ne0$ (
$\Delta \tilde{E}/\tilde{E}< 0.05$, $P>95\%$  empty red circles) for different system sizes in
Fig.~\ref{lmg_central_eig_ent_vs_size_2:fig}. In all cases a logarithmic scaling with
the size is achieved.
%
%\begin{figure}
%\epsfig{file=plot/ent_inf_vs_timeWL16.eps,width=7.2cm,angle=0,clip=}
%\caption{(Color online) Entanglememnt entropy (left panel) and $P_{pos}$ (right panel )as function of the time. 
%The control is effective only for negative time; at $t=0$ the Hamiltonian is kept constant. 
%Different colours correspond to different intensities of of the energy fluctuations in the control cost 
%function; the blue circles represent entropy of the static eigenstates of $H[\Gamma _0]$. The size of 
%the system is N=16.}
%\label{ent_inf_vs_timeWL16_fig}
%\end{figure}
%

\begin{figure}
\centering\epsfig{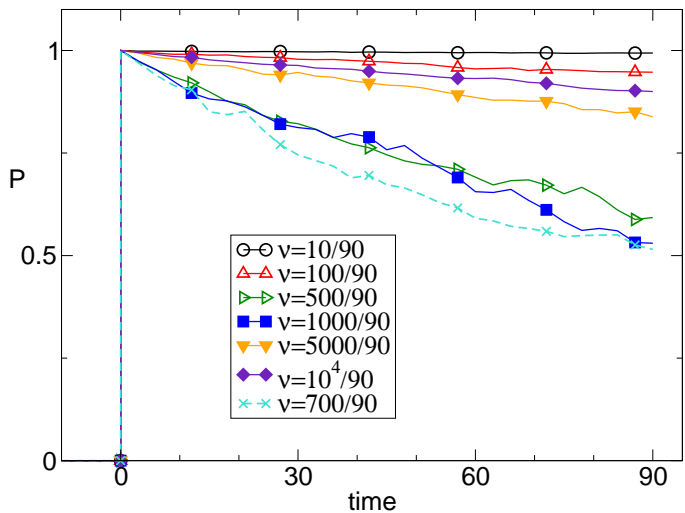}
\caption{(Color online)  Survival probability 
$P(t)$ as a function of time  (time unit $C^{-1}$), averaged over $30$ noise instances for 
$I_{\alpha}=I_{\beta}=0.2$,  $N=64$, $\lambda=1.8$, and different noise
frequencies. The worst case (dashed line with crosses) is for $\nu_R=700/(90C^{-1})= 7.8 C$.}
\label{comparison_inf_vs_timeNL64r0r1.8_1:fig}
\end{figure}
\begin{figure}
\centering\epsfig{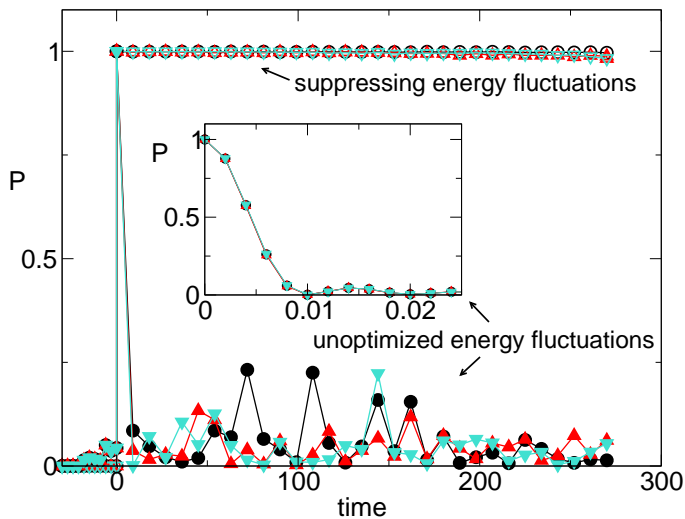}
\caption{(Color online) Survival probability $P(t)$ as a function of
  time (time unit $C^{-1}$) for three realizations of the noise with $I_{\alpha}=I_{\beta}=0.01$ at
  frequency $\nu_R$, %system size 
$N=64$, and $\lambda=1.8$ (empty symbols) or $\lambda=0$ (full symbols). 
Inset: Blow up of the region around $t=0$ for the $\lambda=0$ case.}
\label{comparison_inf_vs_timeNL64r0r1.8_2:fig}
\end{figure}
%

%%%%%%%%%%%%%%%%%%%%%%%%%%
\section{Random telegraph noise}
\label{noise:sec}
%%%%%%%%%%%%%%%%%%%%%%%%%%
A reliable ESU should be robust against external noise and decoherence
even when the control is switched off, in such a way that it could be 
used for subsequent quantum operations.
In order to test the robustness of the optimized states, we model the
effect of decoherence by adding a random telegraph noise and we monitor the time 
evolution in such noisy environment~\cite{Nielsen_Chuang:book}.
In particular we study the evolution induced by the Hamiltonian 
\begin{eqnarray}
H=-\frac{1}{N}[1+I_{\alpha}\alpha (t)]J_x^2-\tilde{\Gamma} [1+I_{\beta}\beta (t)]  J_z
\label{ham_noise:eq}
\end{eqnarray}
where $\alpha (t),\beta (t)$ are random functions of the time with a flat distribution
 in $[-I_j,I_j]$  ($j=\alpha, \beta$), changing random value every typical time $1/\nu$.
The case $I_\alpha= I_\beta=0$ corresponds to a noiseless evolution. The first important observation is that the 
frequency $\nu$ of the signal fluctuations is crucial in determining its effects~\cite{facchi05}. Indeed in
Fig.~\ref{comparison_inf_vs_timeNL64r0r1.8_1:fig}, the survival probability
$P(t)$ is plotted as a function of the time in the presence of a strong noise, $I_\alpha= I_\beta=0.2$,
for a system of $N=64$ spins and for a given initial optimal state obtained with $\lambda=1.8$ 
(see Fig.~\ref{ent_inf_vs_timeWL64_1:fig}). When $\nu$ is either too low 
(empty circles) or too high (full diamonds) the effect of the noise is reduced; however
around a resonant frequency $\nu_R$ (dashed line with crosses) its effect is enhanced and the state is 
quickly destroyed. We checked that the resonant frequency is the same for different eigenvalues, different sizes,
 and different noise strengths (data not shown),
reflecting the fact that in the paramagnetic phase ($\tilde{\Gamma} \gg 1$) 
the gap separating the eigenstates is proportional to $\tilde{\Gamma}$ independently of the size of the system and 
of the state itself, see Eq.~(\ref{ham_lmg:eq}). Therefore we analyze this
worst case scenario, setting $\nu=\nu_R$ from now on. 
%CUT The results of this analysis, show that ESU states -- differently 
%from the states produced optimizing only entanglement -- are extremely robust to noise at the resonant frequency. 
%This is shown in Fig.~\ref{comparison_inf_vs_timeNL64r0r1.8_fig} where
In Fig.~\ref{comparison_inf_vs_timeNL64r0r1.8_2:fig}
%
%\begin{figure}
%\epsfig{file=plot/inf_vs_time_dis0.2ndvarNL64r1.8.eps,width=7cm,angle=0,clip=}
%\caption{(Color online) Survival probability $P(t)$ as function of the time.
%Different colours correspond to different frequencies of the noise; the relative intensity of the noise is fixed
%at a (large) value $0.2$; the data are the average  value of a sample of $30$ instances of disorder. 
%The size of the system is $N=64$ and $r=1.8$.}
%\label{inf_vs_time_dis0.2ndvarNL64r1.8_fig}
%\end{figure}
%
%
\begin{figure}
\centering\epsfig{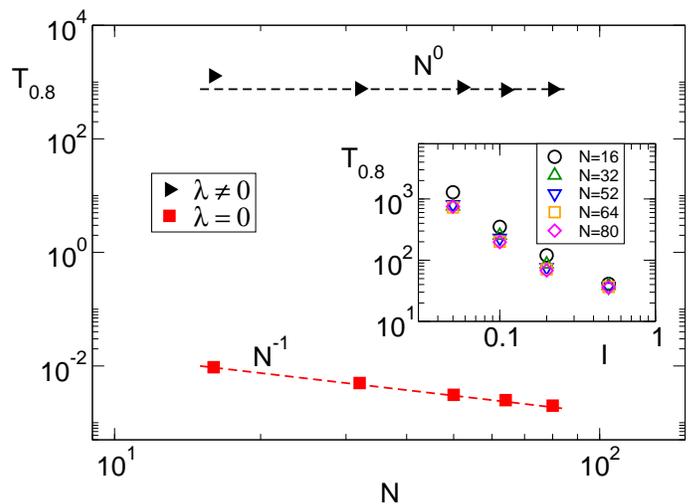}
\caption{(Color online)  
  Time $T_{0.8}$ required to reduce the survival probability $P$ 
  below $0.8$ 
  for different prepared states $|\psi(0)\rangle$ with
  $\lambda=0$ (red squares) and $\lambda \ne 0$ corresponding to the 
  second eigenstate of the even parity sector of $J_z$ (black
  triangles) as a function of the system size $N$. The dashed lines are
  fits of the four rightmost points (biggest system sizes) resulting in $N^{-0.97}$ and $N^{-0.03}$ respectively. Inset: Time $T_{0.8}$ as a function of the intensity 
  $I=I_\alpha= I_\beta$ of the disorder for different system sizes $N$.}
\label{lifetime_vs_I_fig}
\end{figure}
we compare the survival probability $P(t)$ for three instances of the 
disorder at the resonant frequency with an intensity of 
the disorder $I_\alpha= I_\beta=0.01$.
The noise-induce dynamics of the states obtained optimizing 
only with respect to the entanglement 
(i.e. setting $\lambda=0$, full symbols in 
Fig.~\ref{comparison_inf_vs_timeNL64r0r1.8_2:fig}) drastically depends on the
(in general unknown) details of the noise
affecting the system;
%CUT , as shown by the different evolutions 
%induced by different instances of the noise. 
%Thus, such states cannot be used 
%as ESU, unlike those prepared with $\lambda \ne 0$ that 
%are stable, noise-independent long-living entanglement states. 
thus, such states cannot be used as ESU.
Viceversa the states prepared with $\lambda \ne 0$ (empty symbols in 
Fig.~\ref{comparison_inf_vs_timeNL64r0r1.8_2:fig}) 
turn out to be stable, noise-independent, and long-living entanglement.
Finally, in Fig.~\ref{lifetime_vs_I_fig} we study the decay times of 
the survival probability $P(t)$  studying the time $T_{0.8}$ 
needed to drop below a given threshold $P_{min}=0.8$ 
as a function of the system size $N$ and of the 
intensity of the disorder $I=I_\alpha= I_\beta$ (inset). These results
clearly show that $T_{0.8}$ for ESU states is almost independent
from the system size, reflecting the fact that the energy gaps
in this region of the spectrum are mostly size independent. 
Notice that, on the contrary, $T_{0.8}$ for maximally entangled 
states decays linearly with the system size and that there are more than 
four orders of magnitude of difference in the decay times $\tau_\lambda$ 
and  $\tau_0$. 
%The reported values of $T_{0.8}$ for an ESU state 
%--for the experimental implementation of the system
%described by Eq.~(\ref{ham_lmg:eq}) by means of cold atoms in cavities-- 
%would correspond to times of the order of $0.2 \,\mathrm{sec}$ for an 
%interaction of the order of $J \sim 2 \pi \cdot 0.74 kHz$~\cite{Baumann_NAT10}.
Finally, the inset of Fig.~\ref{lifetime_vs_I_fig} shows that  the scaling of $T_{0.8}$
with the noise strength for ESU states is approximately a power law
%linear 
and again depends very weakly on the system size $N$. 
%{\bf This effect is better shown in Fig.~\ref{lifetime_vs_I_fig} where the time required to reduce the 
%survival probability $P(t)$ below a given threshold $P_{min}$, is plotted as a 
%function of the intensity of the disorder $I=I_\alpha= I_\beta$. Different symbols correspond
%to different sizes and different values of $r$. The quite similar behavior of the data
%again supports the scalability of the procedure. 
%In conclusion, the states prepared with $\lambda=0$
%---although characterized by higher entanglement--- cannot be used as a
%stable ESU. On the contrary, the states prepared with $\lambda \ne 0$
%are very robust and insesitive to the noise (empty symbols) even after long times.
%

%%%%%%%%%%%%%%%%%%%%%%%%%%
\section{Ising model: concurrence between extremal spins}
\label{ising:sec}
%%%%%%%%%%%%%%%%%%%%%%%%%%

In our previous discussion we focused our attention onto the optimization of the Von Neumann
entropy of eigenstates other than the ground state of the LMG model, in order to
show the effectiveness of our protocol in controlling the dynamics and unexplored 
properties of many-body systems.
\begin{figure}
\centering\epsfig{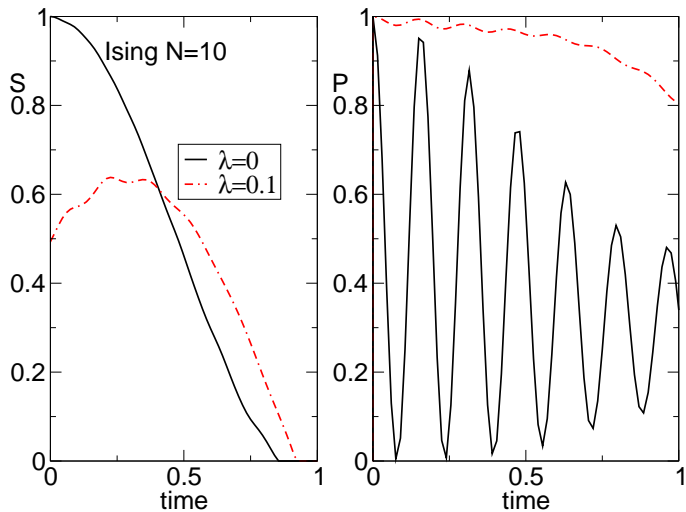}
\caption{(Color online)  
  Concurrence $S(t)$ between extremal spins (left panel) and survival probability $P(t)$ 
  (right panel) as 
  a function of time (time units $C^{-1}$) in an Ising chain with $N=10$ spins for two different
  $\lambda$ values: $\lambda =0$ black continuous line and $\lambda =0.1$ red dot-dashed line.}
\label{ising_N10_fig}
\end{figure}
However aiming at demonstrating the generality of the method, in this section we would like 
to present briefly the application of our protocol to a different 
situation, closer to the typical problems encountered in quantum information:
in particular we are showing how it is possible to stabilize the concurrence between 
the extremal spins of an open Ising chain.\\
The Hamiltonian of the ordered one-dimensional Ising model with nearest neighbor 
interaction is given by:
\begin{eqnarray}
H=-C\sum ^{N-1} _{i}\sigma _i ^x \sigma _{i+1} ^x -
    \Gamma (t) \sum _{i}^N \sigma _i ^z ,
\label{full_ham_ising:eq}
\end{eqnarray}
where the transverse field $\Gamma (t)$ is our control field. We assume that the system
can be easily prepared in the ground state at a large value of the control field 
$\tilde{\Gamma}=10$, in which all the spins are polarized along the positive $z$-direction.
The aim of the control is to enhance the concurrence between the 
first and the $N$-th spin of the chain, possibly stabilizing the state.
The concurrence between two spins is defined as
$S={\rm max}\{0,e_1-e_2-e_3-e_4\}$, where the $e_i$'s are the eigenvalues in decreasing
order of the Hermitian matrix $R=\sqrt{\sqrt{\rho}\tilde{\rho}\sqrt{\rho}}$,
$\rho$ is the reduced density matrix of the two extremal spins, and
$\tilde{\rho}=(\sigma _y \otimes\sigma _y)\rho ^*(\sigma _y \otimes\sigma _y)$ 
is the spin-flipped state~\cite{Wootters_PRL98}.\\ 
At a large value of the transverse field, 
the eigenstates of the Hamiltonian are the classical states represented 
by all the possible up-down combinations of $N$ spins, and states with the same numbers of flipped
spins, though in different positions, are degenerate. A naive approach to build stable
entangled states would then require a search
for possibly entangled states in each degenerate subspace at a given energy. Such a search
however represents a highly non trivial task, due to the strong constraint imposed by requiring
non vanishing concurrence: again a suitable recipe for such a search should be provided
and is non-trivial to find. 
On the contrary our protocol proposes an answer to the task without requiring any 
diagonalization, while automatically performing the search, therefore offering a clear advantage.\\
We perform a CRAB optimization in the time interval $[-T,0]$ minimizing the function
$\mathcal{F}(\lambda)_{|\psi (0)\rangle}=-S+ \lambda \Delta \tilde{E}$,
in which now $S$ is the concurrence; then at the time $t=0$ the control is
switched off, the value of the field is kept constant ($\Gamma(t)=\tilde{\Gamma}$ for
$t>0$), and we observe the evolution of the optimized state.
In Fig.~\ref{ising_N10_fig} we show the behavior of the concurrence $S(t)$ and of the
survival probability $P(t)=\left({\rm Tr}\sqrt{\sqrt{\rho (t)}\rho (0)\sqrt{\rho (t)}}\right)^2$,
excluding ($\lambda=0$ black continuous line) and including ($\lambda=0.1$ red dot-dashed line) 
the energy fluctuation term in the optimization procedure. As shown in the picture,
although, as expected, the concurrence is smaller 
when $\lambda\neq 0$, the survival probability is stabilized by a factor bigger than $50$
in time with respect to the $\lambda=0$ case.

%%%%%%%%%%%%%%%%%%%%%%%%%%
\section{Conclusions}
\label{conclusion:sec}
%%%%%%%%%%%%%%%%%%%%%%%%%%

Exploiting optimal control we proposed a method to steer a system into 
apriori unknown eigenstates satisfying desired properties. We demonstrated,
on a particular system, that this protocol can be effectively used to build
long-lived entangled states with many-body systems, indicating a
possible implementations of an Entanglement Storage Unit 
scalable with the system size. 
The presented method is compatible 
with different models (e.g. LMG and Ising) and 
measures of entanglement (e.g. von Neumann entropy and concurrence)
and it can be extended to any other property one is interested in, as for 
example the squeezing of the target state~\cite{inpreparation}. 
It can be applied to different systems with apriori unknown
properties: optimal control will select the states (if any) satisfying the
desired property and robust to system perturbations. We stress that an 
adiabatic strategy is absolutely ineffective for this purpose, as
transitions between different eigenstates are forbidden. 
Applying this protocol to the full open-dynamics description of the
system, e.g. via a CRAB optimization of the Lindblad dynamics as done
in~\cite{caruso11}, will result in an optimal search 
of a Decoherence Free Subspace (DFS) with desired
properties~\cite{DFS}. If no DFS exists, the optimization would lead
the system in an eigenstate of the superoperator with longest
lifetime and desired properties~\cite{inpreparation}. Although the
state so prepared may be unstable over long times, it represents 
the best and most robust state attainable, and additional 
(weak) control might be used to preserve its stability. 
%Combining this
%approach with the recently introduced enviroment enginnering might
%lead to even better results~\cite{kraus}.
Finally, working with excited states would reduce finite
temperature effects, relaxing low temperatures 
working-point conditions, simplifying the experimental requirements
to build a reliable ESU.

%\ack

We acknowledge discussions with M.~D. Lukin, and support from the EU projects AQUTE, PICC, the
SFB/TRR21 and the BWgrid for computational resources.

%%%%%%%%%%%%%%%%%%%%%%%%%%%%%%%%%%%%%%%%%%%%%%%%%%%%%%%%%%%%%%%%%%%%%%%%
%                               BIBLIOGRAPHY             %%%%%%%%%%%%%%%
%%%%%%%%%%%%%%%%%%%%%%%%%%%%%%%%%%%%%%%%%%%%%%%%%%%%%%%%%%%%%%%%%%%%%%%%

%\section*{References}

%\bibliography{QA}

\begin{thebibliography}{40}

\bibitem{Nielsen_Chuang:book} Nielsen~M and Chuang I~L 2000 {\it Quantum Computation and Quantum Information} 
(Cambridge University Press)
%\bibitem{Nielsen_Chuang:book} M.~Nielsen and I.~L.~Chuang, \emph{Quantum Computation and Quantum Information} 
%(Cambridge University Press, 2000)
%
% \bibitem[{\citenamefont{Nielsen and Chuang}(2000)}]{Nielsen_Chuang:book}
% \bibinfo{author}{\bibfnamefont{M.}~\bibnamefont{Nielsen}} \bibnamefont{and}
%   \bibinfo{author}{\bibfnamefont{I.~L.} \bibnamefont{Chuang}},
%   \emph{\bibinfo{title}{Quantum Computation and Quantum Information}}
%   (\bibinfo{publisher}{Cambridge University Press}, \bibinfo{year}{2000}).

\bibitem{Baumann_NAT10} Baumann K, Guerlin C, Brennecke F and Esslinger T 2010  {\it Nature} {\bf 464} 1301 
%\bibitem{Baumann_NAT10} K.~Baumann, C.~Guerlin, F.~Brennecke, and T.~Esslinger,  Nat. {\bf 464}, 1301 (2010)
%
% \bibitem[{\citenamefont{Baumann et~al.}(2010)\citenamefont{Baumann, Guerlin,
%   Brennecke, and Esslinger}}]{Baumann_NAT10}
% \bibinfo{author}{\bibfnamefont{K.}~\bibnamefont{Baumann}},
%   \bibinfo{author}{\bibfnamefont{C.}~\bibnamefont{Guerlin}},
%   \bibinfo{author}{\bibfnamefont{F.}~\bibnamefont{Brennecke}},
%   \bibnamefont{and}
%   \bibinfo{author}{\bibfnamefont{T.}~\bibnamefont{Esslinger}},
%   \bibinfo{journal}{Nat.} \textbf{\bibinfo{volume}{464}}, \bibinfo{pages}{1301}
%   (\bibinfo{year}{2010}).

\bibitem{Amico_RMP08} Amico L, Fazio R, Osterloh A and Vedral V 2008 {\it Rev. Mod. Phys.} {\bf 80} 517 
%\bibitem{Amico_RMP08} L.~Amico, R.~Fazio, A.~Osterloh, and V.~Vedral, Rev. Mod. Phys. {\bf 80}, 517 (2008)
%
% \bibitem[{\citenamefont{Amico et~al.}(2008)\citenamefont{Amico, Fazio,
%   Osterloh, and Vedral}}]{Amico_RMP08}
% \bibinfo{author}{\bibfnamefont{L.}~\bibnamefont{Amico}},
%   \bibinfo{author}{\bibfnamefont{R.}~\bibnamefont{Fazio}},
%   \bibinfo{author}{\bibfnamefont{A.}~\bibnamefont{Osterloh}}, \bibnamefont{and}
%   \bibinfo{author}{\bibfnamefont{V.}~\bibnamefont{Vedral}},
%   \bibinfo{journal}{Rev. Mod. Phys.} \textbf{\bibinfo{volume}{80}},
%   \bibinfo{pages}{517} (\bibinfo{year}{2008}).

\bibitem{Vidal1_PRL03} Vidal G, Latorre J, Rico E and Kitaev A 2003 {\it Phys. Rev. Lett.} {\bf 90} 227902
%\bibitem{Vidal1_PRL03} G.~Vidal, J.~Latorre, E.~Rico, and A.~Kitaev, Phys. Rev. Lett. {\bf 90}, 227902 (2003)
%
% \bibitem[{\citenamefont{Vidal et~al.}(2003)\citenamefont{Vidal, Latorre, Rico,
%   and Kitaev}}]{Vidal1_PRL03}
% \bibinfo{author}{\bibfnamefont{G.}~\bibnamefont{Vidal}},
%   \bibinfo{author}{\bibfnamefont{J.}~\bibnamefont{Latorre}},
%   \bibinfo{author}{\bibfnamefont{E.}~\bibnamefont{Rico}}, \bibnamefont{and}
%   \bibinfo{author}{\bibfnamefont{A.}~\bibnamefont{Kitaev}},
%   \bibinfo{journal}{Phys. Rev. Lett.} \textbf{\bibinfo{volume}{90}},
%   \bibinfo{pages}{227902} (\bibinfo{year}{2003}).

\bibitem{Alba_JSTAT09} Alba V, Fagotti M and Calabrese P 2009 {\it J. Stat. Mech.} p. P10020
%\bibitem{Alba_JSTAT09} V.~Alba, M.~Fagotti, and P.~Calabrese, J. Stat. Mech. p. P10020 (2009)
%
%\bibitem[{\citenamefont{Alba et~al.}(2009)\citenamefont{Alba, Fagotti, and
%   Calabrese}}]{Alba_JSTAT09}
% \bibinfo{author}{\bibfnamefont{V.}~\bibnamefont{Alba}},
%   \bibinfo{author}{\bibfnamefont{M.}~\bibnamefont{Fagotti}}, \bibnamefont{and}
%   \bibinfo{author}{\bibfnamefont{P.}~\bibnamefont{Calabrese}},
%   \bibinfo{journal}{J. Stat. Mech.} p. \bibinfo{pages}{P10020}
%   (\bibinfo{year}{2009}).

\bibitem{Alcaraz_PRL11} Alcaraz F C, Berganza M and Sierra G 2011 {\it Phys. Rev. Lett.} {\bf 106} 201601
%\bibitem{Alcaraz_PRL11} F.~C.~Alcaraz, M.~Berganza, and G.~Sierra, Phys. Rev. Lett. {\bf 106}, 201601 (2011)
%
%\bibitem[{\citenamefont{Alcaraz et~al.}(2011)\citenamefont{Alcaraz, Berganza,
%  and Sierra}}]{Alcaraz_PRL11}
% \bibinfo{author}{\bibfnamefont{F.~C.} \bibnamefont{Alcaraz}},
%   \bibinfo{author}{\bibfnamefont{M.}~\bibnamefont{Berganza}}, \bibnamefont{and}
%   \bibinfo{author}{\bibfnamefont{G.}~\bibnamefont{Sierra}},
%   \bibinfo{journal}{Phys. Rev. Lett.} \textbf{\bibinfo{volume}{106}},
%   \bibinfo{pages}{201601} (\bibinfo{year}{2011}).

\bibitem{Doria_PRL11} Doria P, Calarco T and Montangero S 2011 {\it Phys. Rev. Lett.} {\bf 106} 190501
%\bibitem{Doria_PRL11} P~Doria, T.~Calarco, and S.~Montangero, Phys. Rev. Lett. {\bf 106},190501 (2011)
%
% \bibitem[{\citenamefont{Doria et~al.}(2011)\citenamefont{Doria, Calarco, and
%   Montangero}}]{Doria_PRL11}
% \bibinfo{author}{\bibfnamefont{P.}~\bibnamefont{Doria}},
%   \bibinfo{author}{\bibfnamefont{T.}~\bibnamefont{Calarco}}, \bibnamefont{and}
%   \bibinfo{author}{\bibfnamefont{S.}~\bibnamefont{Montangero}},
%   \bibinfo{journal}{Phys. Rev. Lett.} \textbf{\bibinfo{volume}{106}},
%   \bibinfo{pages}{190501} (\bibinfo{year}{2011})

\bibitem{Caneva_PRA11} Caneva T, Calarco T and Montangero S 2011 {\it Phys. Rev. A} {\bf 84} 022326
%\bibitem{Caneva_PRA11} T.~Caneva, T.~Calarco, and S.~Montangero, Phys. Rev. A {\bf 84}, 022326 (2011)

\bibitem{Platzer_PRL10} Platzer F, Mintert F and Buchleitner A 2010 {\it Phys. Rev. Lett.} {\bf 105} 020501
%\bibitem{Platzer_PRL10} F.~Platzer, F.~Mintert, and A.~Buchleitner, Phys. Rev. Lett. {\bf 105}, 020501 (2010)
%
% \bibitem[{\citenamefont{Platzer et~al.}(2010)\citenamefont{Platzer, Mintert,
%   and Buchleitner}}]{Platzer_PRL10}
% \bibinfo{author}{\bibfnamefont{F.}~\bibnamefont{Platzer}},
%   \bibinfo{author}{\bibfnamefont{F.}~\bibnamefont{Mintert}}, \bibnamefont{and}
%   \bibinfo{author}{\bibfnamefont{A.}~\bibnamefont{Buchleitner}},
%   \bibinfo{journal}{Phys. Rev. Lett.} \textbf{\bibinfo{volume}{105}},
%   \bibinfo{pages}{020501} (\bibinfo{year}{2010}).

\bibitem{Lipkin_NP65} Lipkin H J, Meshkov N and Glick A J 1965 {\it Nucl. Phys.} {\bf 62} 188
%\bibitem{Lipkin_NP65} H.~J.~Lipkin, N.~Meshkov, A.~J.~Glick, Nucl. Phys. {\bf 62}, 188 (1965)
%
% \bibitem[{\citenamefont{Lipkin et~al.}(1965)\citenamefont{Lipkin, Meshkov, and
%   Glick}}]{Lipkin_NP65}
% \bibinfo{author}{\bibfnamefont{H.~J.} \bibnamefont{Lipkin}},
%   \bibinfo{author}{\bibfnamefont{N.}~\bibnamefont{Meshkov}}, \bibnamefont{and}
%   \bibinfo{author}{\bibfnamefont{A.~J.} \bibnamefont{Glick}},
%   \bibinfo{journal}{Nucl. Phys.} \textbf{\bibinfo{volume}{62}},
%   \bibinfo{pages}{188} (\bibinfo{year}{1965}).

\bibitem{Buecker_NAT11} B\"ucker R, Grond J, Manz S, Berrada T, Betz T, Koller C, Hohenester U,
Schumm T, Perrin A and Schmiedmayer J 2011 {\it Nat. Phys.} {\bf 7} 608 
%\bibitem{Buecker_NAT11} R.~B\"ucker, J.~Grond, S.~Manz, T.~Berrada, T.~Betz, C.~Koller, U.~Hohenester,
%T.~Schumm, A.~Perrin, and J.~Schmiedmayer, Nat. Phys. {\bf 7}, 608 (2011)
%
% \bibitem[{\citenamefont{Bucker et~al.}(2011)\citenamefont{Bucker, Grond, Manz,
%   Berrada, Betz, Koller, Hohenester, Schumm, Perrin, and
%   Schmiedmayer}}]{Buecker_NAT11}
% \bibinfo{author}{\bibfnamefont{R.}~\bibnamefont{Bucker}},
%   \bibinfo{author}{\bibfnamefont{J.}~\bibnamefont{Grond}},
%   \bibinfo{author}{\bibfnamefont{S.}~\bibnamefont{Manz}},
%   \bibinfo{author}{\bibfnamefont{T.}~\bibnamefont{Berrada}},
%   \bibinfo{author}{\bibfnamefont{T.}~\bibnamefont{Betz}},
%   \bibinfo{author}{\bibfnamefont{C.}~\bibnamefont{Koller}},
%   \bibinfo{author}{\bibfnamefont{U.}~\bibnamefont{Hohenester}},
%   \bibinfo{author}{\bibfnamefont{T.}~\bibnamefont{Schumm}},
%   \bibinfo{author}{\bibfnamefont{A.}~\bibnamefont{Perrin}}, \bibnamefont{and}
%   \bibinfo{author}{\bibfnamefont{J.}~\bibnamefont{Schmiedmayer}},
%   \bibinfo{journal}{Nat. Phys.}  (\bibinfo{year}{2011}).

\bibitem{inpreparation} Caneva T et al {\it in preparation}

\bibitem{Anandan_PRL90} Anandan J and Aharonov Y 1990 {\it Phys. Rev. Lett.} {\bf 65} 1697 

\bibitem{Botet_PRB83} Botet R and Jullien R 1983 {\it Phys. Rev.} {\bf 28} 3955 
%\bibitem{Botet_PRB83} R. Botet and R. Jullien, Phys. Rev. {\bf 28}, 3955 (1983)

\bibitem{Latorre_JPA09} Latorre J I and Riera A 2009 {\it J. Phys. A: Math. Theor.} {\bf 42} 504002
%\bibitem{Latorre_JPA09} J.~I.~Latorre and A.~Riera, J. Phys. A: Math. Theor. {\bf 42}, 504002 (2009)
%
% \bibitem[{\citenamefont{Latorre and Riera}(2009)}]{Latorre_JPA09}
% \bibinfo{author}{\bibfnamefont{J.~I.} \bibnamefont{Latorre}} \bibnamefont{and}
%   \bibinfo{author}{\bibfnamefont{A.}~\bibnamefont{Riera}}, \bibinfo{journal}{J.
%   Phys. A: Math. Theor.} \textbf{\bibinfo{volume}{42}}, \bibinfo{pages}{504002}
%   (\bibinfo{year}{2009}).

%new
\bibitem{Latorre_PRA05} Latorre J I, Orus R, Rico E and Vidal J 2005 {\it Phys. Rev. A} {\bf 71} 064101 \\ 
Barthel T, Dusuel S and Vidal J 2006 {\it Phys. Rev. Lett.} {\bf 97} 220402 \\ 
For the solution of the LMG model in the thermodynamical limit see also: \\
Ribeiro P, Vidal J and Mosseri R 2008 {\it Phys. Rev. E} {\bf 78} 021106
%%%%%%%%%%%%%%%%%%%%%%%%%%%%%%%%%%%%%%%%%%%%%%%%%%%%%%%%%%%%%%%%%%%%%%%%%%%%%%%
%old
% \bibitem{Latorre_PRA05} J.~I.~Latorre, R.~Orus, E.~Rico, and J.~Vidal, Phys. Rev. A {\bf 71} 064101 (2005);
% V. Popkov and M. Salerno, Phys. Rev. A {\bf 71} 012301 (2005). {\bf For the solution
% of the LMG model in the thermodynamical limit see also P~Ribeiro, J.~Vidal, and R.~Mosseri
% Phys. Rev. E {\bf 78}, 021106 (2008)}.
%
% \bibitem[{\citenamefont{Latorre et~al.}(2005)\citenamefont{Latorre, Orus, Rico,
%   and Vidal}}]{Latorre_PRA05}
% \bibinfo{author}{\bibfnamefont{J.~I.} \bibnamefont{Latorre}},
%   \bibinfo{author}{\bibfnamefont{R.}~\bibnamefont{Orus}},
%   \bibinfo{author}{\bibfnamefont{E.}~\bibnamefont{Rico}}, \bibnamefont{and}
%   \bibinfo{author}{\bibfnamefont{J.}~\bibnamefont{Vidal}},
%   \bibinfo{journal}{Phys. Rev. A} \textbf{\bibinfo{volume}{71}},
%   \bibinfo{pages}{064101} (\bibinfo{year}{2005});
% V. Popkov and M. Salerno, Phys. Rev. A {\bf 71} 012301 (2005). 
% {\bf For the solution
% of the LMG model in the thermodynamical limit see also P~Ribeiro, J.~Vidal, and R.~Mosseri
% Phys. Rev. E {\bf 78}, 021106 (2008)}.

% old
% \bibitem{Barthel_PRL06} T.~Barthel, S.~Dusuel, and J.~Vidal, Phys. Rev. Lett. {\bf 97}, 220402 (2006);
% J.~Vidal, S.~Dusuel, T.~Barthel, J. Stat. Mech. p. {P01015} (2007).
%
% \bibitem[{\citenamefont{Barthel et~al.}(2006)\citenamefont{Barthel, Dusuel, and
%   Vidal}}]{Barthel_PRL06}
% \bibinfo{author}{\bibfnamefont{T.}~\bibnamefont{Barthel}},
%   \bibinfo{author}{\bibfnamefont{S.}~\bibnamefont{Dusuel}}, \bibnamefont{and}
%   \bibinfo{author}{\bibfnamefont{J.}~\bibnamefont{Vidal}},
%   \bibinfo{journal}{Phys. Rev. Lett.} \textbf{\bibinfo{volume}{97}},
%   \bibinfo{pages}{220402} (\bibinfo{year}{2006});
%   \bibinfo{author}{\bibfnamefont{J.}~\bibnamefont{Vidal}},
%   \bibinfo{author}{\bibfnamefont{S.}~\bibnamefont{Dusuel}}, \bibnamefont{and}
%   \bibinfo{author}{\bibfnamefont{T.}~\bibnamefont{Barthel}},
%   \bibinfo{journal}{J. Stat. Mech.} p. \bibinfo{pages}{P01015}
%   (\bibinfo{year}{2007}).
%%%%%%%%%%%%%%%%%%%%%%%%%%%%%%%%%%%%%%%%%%%%%%%%%%%%%%%%%%%%%%%%%%%%%%%%%%%%%%%%%%%%%%%%%%%%%%

\bibitem{facchi05} Facchi P, Montangero S, Fazio R and Pascazio P 2005 {\it Phys. Rev. A} {\bf 71} 060306 

\bibitem{Wootters_PRL98} Wootters W K 1998 {\it Phys. Rev. Lett.} {\bf 80} 2245

\bibitem{caruso11}  Caruso F, Montangero S, Calarco T, Huelga S F and Plenio M B 2011 {\it Preprint} arXiv:1103.0929

\bibitem{DFS} Palma G M, Suominen K A and Ekert A K 1996 {\it Proc. Roy. Soc. Lond. A} {\bf 452} 567 \\ 
Duan L M and Guo G C 1997 {\it Phys. Rev. Lett.} {\bf 79} 1953 \\
Zanardi P and Rasetti M 1997 {\it Phys. Rev. Lett.} {\bf 79} 3306 


\end{thebibliography}
\bibliographystyle{apsrev}

\end{document}